\documentclass[plb,preprint,tightenlines,floatfix,showpacs,preprintnumbers,nofootinbib]{revtex4}
 \usepackage[dvips,final]{graphicx}
  \usepackage{amssymb}
   \usepackage{amsmath}
    \usepackage{epsfig}
     \usepackage{bm}
      \usepackage{pifont}
\usepackage{xcolor}
\def\Li{\relax\ifmmode{\textbf{Li}_{2}}\else{Li$_2${ }}\fi}

\newcommand{\be}{\begin{equation}}
\newcommand{\ee}{\end{equation}}

\newcommand{\ba}{\begin{eqnarray}}
\newcommand{\ea}{\end{eqnarray}}
\newcommand{\bg}{\begin{gather}}
\newcommand{\foma}{\end{gather}}

\newcommand{\vecc}[1]{\mbox{\boldmath $#1$}}

\def\<{\langle}
\def\>{\rangle}

\def\({\left(}
\def\[{\left[}
\def\){\right)}
\def\]{\right]}

\begin{document}
\thispagestyle{empty}

\title{Cusped light-like Wilson loops in gauge theories}
\author{I.O.~Cherednikov}
\email{igor.cherednikov@ua.ac.be}
\affiliation{Departement Fysica, Universiteit Antwerpen, B-2020 Antwerpen, Belgium\\
and\\
BLTP JINR, RU-141980 Dubna, Russia}
\author{T.~Mertens}
\email{tom.mertens@ua.ac.be}
\affiliation{Departement Fysica, Universiteit Antwerpen, B-2020 Antwerpen, Belgium\\}
\author{F.F.~Van der Veken}
\email{frederik.vanderveken@ua.ac.be}
\affiliation{Departement Fysica, Universiteit Antwerpen, B-2020 Antwerpen, Belgium\\}
\vspace {10mm}
\date{\today}

\begin{abstract}
We propose and discuss a new approach to the analysis of the correlation functions which contain light-like Wilson lines or loops, the latter being cusped in addition. The objects of interest are therefore the light-like Wilson null-polygons, the soft factors of the parton distribution and fragmentation functions, high-energy scattering amplitudes in the eikonal approximation, gravitational Wilson lines, etc. Our method is based on a generalization of the universal quantum dynamical principle by J. Schwinger and allows one to take care of extra singularities emerging due to light-like or semi-light-like cusps. We show that such Wilson loops obey a differential equation which connects the area variations and renormalization group behavior of those objects and discuss the possible relation between geometrical structure of the loop space and area evolution of the light-like cusped Wilson loops.
\end{abstract}
\maketitle



\section{Introduction}
\label{sec:intro}
Wilson lines (also known as gauge links or eikonal lines) {can} be naturally introduced in any gauge field theory. These objects are generically defined via traces of path-ordered exponentials of a gauge field evaluated along a given trajectory $ {\cal W} (\Gamma)
=
  {\cal P} \exp
  \left[\ -ig\int_{[{\Gamma}]}\ dz^{\mu} \ {\cal A}_{\mu}(z) \
  \right]
  $.
The path $\Gamma$ is a curve along which the gauge field ${\cal A}$ gets transported from the initial point to the final one.
Wilson lines defined on closed contours are called Wilson loops. They are path-dependent non-local functionals of the gauge field, invariant under gauge group transformations. Putting the matter of question more mathematical, one can construct a space with its elements being Wilson loops defined on an infinite set of contours.
Reformulation of QCD in terms of the elements of a generic loop space would allow one to use gauge-invariant quantities as fundamental degrees of freedom instead of the quarks and gluons from the standard QCD Lagrangian \cite{Loop_Space, WL_RG}. Observables can then be obtained via correlation functions of Wilson loops:
\begin{equation}
 {\cal W}_n (\Gamma_1, ... \Gamma_n)
  = \Big \langle 0 \Big| {\cal T} \frac{1}{N_c} {\rm Tr}\ \Phi (\Gamma_1)\cdot \cdot \cdot \frac{1}{N_c}{\rm Tr}\ \Phi (\Gamma_n)  \Big| 0 \Big\rangle \ , \
  \Phi (\Gamma_i)
   =
   {\cal P} \ \exp\[ig \oint_{\Gamma_i} \ dz^\mu A_{\mu} (z) \] \ .
   \label{eq:wl_def}
\end{equation}
Complete information on the quantum dynamical properties of the loop space is accumulated in the Schwinger-Dyson equations:
\begin{equation}
 \langle 0 | \nabla_\mu F^{\mu\nu} \ {\cal O} (A) \ | 0 \rangle
=
i \langle 0 |  \frac{\delta }{\delta A_\nu} \  {\cal O} (A) \ | 0 \rangle \ ,
\label{eq:sch_dy_YM}
\end{equation}
where ${\cal O} (A)$ stands for an arbitrary functional of the gauge fields. Let the functionals ${\cal O} (A)$ be the Wilson exponentials $\Phi (\Gamma)$ (\ref{eq:wl_def}). Then Eqs. (\ref{eq:sch_dy_YM}) turn into
the Makeenko-Migdal (MM) equations \cite{MM_WL}:
\begin{equation}
 \partial_x^\nu \ \frac{\delta}{\delta \sigma_{\mu\nu} (x)} \ {\cal W}_1(\Gamma)
 =
 N_c g^2 \ \oint_{\Gamma} \ dz^\mu \ \delta^{(4)} (x - z) {\cal W}_2(\Gamma_{xz} \Gamma_{zx}) \ ,
 \label{eq:MM_general}
\end{equation}
where the basic operations are the area- $\delta/\delta\sigma_{\mu\nu}$ and the path- $\partial_\mu$ derivatives \cite{MM_WL}:
\begin{equation}
 \frac{\delta}{\delta \sigma_{\mu\nu} (x)} \ \Phi (\Gamma)
 \equiv
 \lim_{|\delta \sigma_{\mu\nu} (x)| \to 0} \ \frac{ \Phi (\Gamma\delta \Gamma) - \Phi (\Gamma) } {|\delta \sigma_{\mu\nu} (x)|} \ ,
\label{eq:area_derivative}
\end{equation}
and the contour $\Gamma\delta \Gamma$ is obtained from the initial one by means of the infinitesimal area deformation $\delta \Gamma$ at the point $x$, while the path variation without changing the area gives rise to the path derivative
\begin{equation}
 \partial_\mu  \Phi(\Gamma)
 =
 \lim_{|\delta x_{\mu}|} \frac{\Phi(\delta x_\mu^{-1}\Gamma\delta x_\mu) - \Phi(\Gamma)}{|\delta x_{\mu}|} \ .
 \label{eq:path_derivative}
\end{equation}
The area derivative can be written as well in the so-called Polyakov form---see, e.g., \cite{St_Kr_WL_cast} for a discussion of an alternative approach.

\begin{figure}[ht]
 $$\includegraphics[angle=90,scale=0.7]{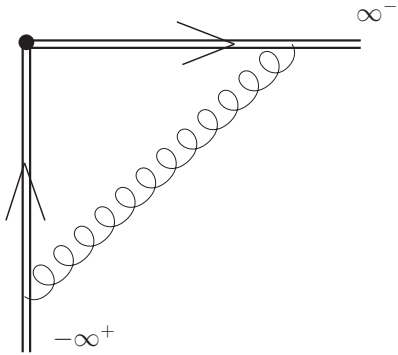}$$
   \vspace{0.0cm}
   \caption{\label{fig:0}The cusped integration contour on the light-cone with the one-gluon exchanges giving rise to the cusp anomalous dimension.}
\end{figure}

Note that the derivation of the MM equations from the Schwinger-Dyson equations is grounded on the Mandelstam formula
\begin{equation}
 \frac{\delta}{\delta \sigma_{\mu\nu} (x)} \ \Phi (\Gamma)
 =
 ig {\rm Tr} \[ F_{\mu\nu} \ \Phi (\Gamma_x)  \]
\end{equation}
and/or on the Stokes theorem, so that the Wilson functionals which do not satisfy the corresponding restrictions (such as, e.g., cusped light-like loops) apparently cannot be straightforwardly treated within the same scheme. There are several other issues limiting the predictive power of the MM equations.  Namely, there exists an interesting class of Wilson loops which possess very specific singularities originating, in particular, from the cusps and/or self-intersections of the contours and, in addition, from the light-like segments of the integration paths. The simplest example is given by a Wilson exponential evaluated along a cusped contour with two semi-infinite light-like sides, Fig. \ref{fig:0}.
Already the leading order contribution to this Wilson exponential possesses all the peculiar singularities: the pure ultraviolet, the infrared (due to the infinite lengths of the sides), and the light-like cups divergences. This simple contour will arise in what follows as a building unit of many important Wilson loops and correlation functions. Physically it corresponds to the soft part of the factorized quark form factor, which has been studied in detail in \cite{WL_LC_rect, CAD_universal}.

In the present work we propose and discuss a new approach to these issues, having in mind, as an instructive example, a very special type of Wilson loops---{planar rectangles} with light-like sides. Considerable interest to cusped light-like Wilson polygons has arisen thanks to the recently conjectured duality between the $n-$gluon planar scattering amplitudes in the ${\cal N} = 4 $ super-Yang-Mills theory and the vacuum average of {planar} Wilson loops formed, correspondingly, by $n$ light-like segments connecting space-time points ${x_i}$, so that their ``lengths'' $x_i - x_{i+1} = p_i$ are chosen equal to the external momenta of the $n-$gluon amplitude (see, e.g., \cite{WL_CFT} and references therein). It has been demonstrated that the infrared singularities of the former corresponds to the ultraviolet singularities of the latter, and the cusp anomalous dimension is the crucial constituent of the evolution equations \cite{KR87}.

Wilson exponentials possessing light-like segments (or that are fully light-like) have been studied also in a different context \cite{WL_LC_rect}. The main observation is that the renormalization properties of these Wilson loops are more intricate than those of cusped Wilson loops defined on off-light-cone integration contours. Namely, the light-cone cusped Wilson loops are not multiplicatively renormalizable because of the additional light-cone singularities (besides the standard ultraviolet and infrared ones). It is possible, however, to construct a combined renormalization-group equation taking into account ultraviolet as well as light-cone divergences. The cusp anomalous dimension, which is the principal ingredient of this equation, is remarkably universal: it controls, e.g., the infrared asymptotic behavior of such important quantities as the QCD and QED Sudakov form factors, the gluon Regge trajectory, the integrated (collinear) parton distribution functions at large-$x$, the anomalous dimension of the heavy quark effective theory, etc. \cite{WL_LC_rect, KR87, CAD_universal, St_Kr_Pheno}.

Another interesting field of application of cusped light-cone Wilson lines could be transverse-momentum dependent parton densities (TMDs) \cite{TMD_singular, CS_all}. The latter are {introduced} to describe the intrinsic transverse momentum of partons inside the nucleon, which is needed in the study of semi-inclusive processes within the (generalization of) the QCD factorization formalism \cite{TMD_singular, TMD_fact}.

\section{Example: singularity structure of TMDs}

\begin{figure}[p]
\includegraphics[width=0.45\textwidth,height=0.70\textheight,angle=90]{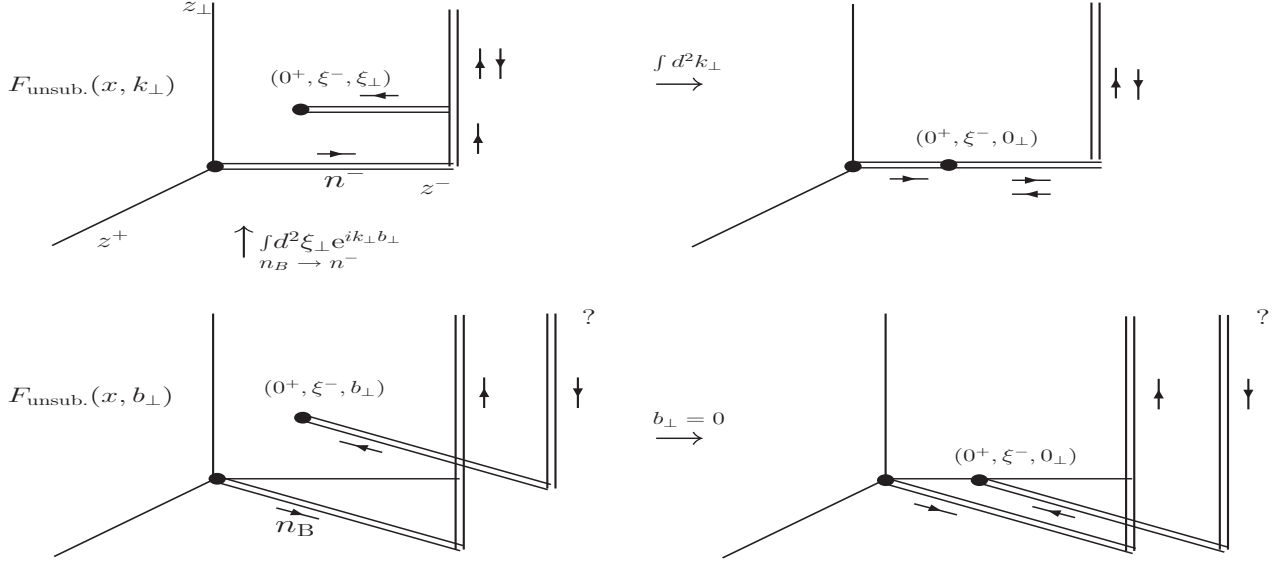}\hspace{2pc}%
\caption{\label{fig:geo1} Geometry of the contours in unsubtracted TMDs with light-like (upper panel) and off-light-cone (lower panel) longitudinal Wilson lines and their reduction to integrated PDFs in tree approximation. In the former case, the transverse Wilson lines vanish after $\bm k_\perp$-integration, while the longitudinal Wilson lines turn into an one-dimensional connector ${\cal W}_n(\xi^-, 0^-)$. In the off-light-cone schemes, the mutual compensation of transverse Wilson lines at infinity is not visible. Moreover, the integrated configuration contains two non-vanishing off-light-cone Wilson lines, which apparently are not equivalent to the collinear connector ${\cal W}_n(\xi^-, 0^-)$. The interrogation marks next to the transverse Wilson lines symbolize the lacking of any consistent treatment in TMD formulations with off-light-cone (shifted) Wilson lines. In contrast, the transverse Wilson lines appear naturally in ``light-cone'' schemes.}
\end{figure}

\begin{figure}[p]
\includegraphics[width=0.45\textwidth,height=0.70\textheight,angle=90]{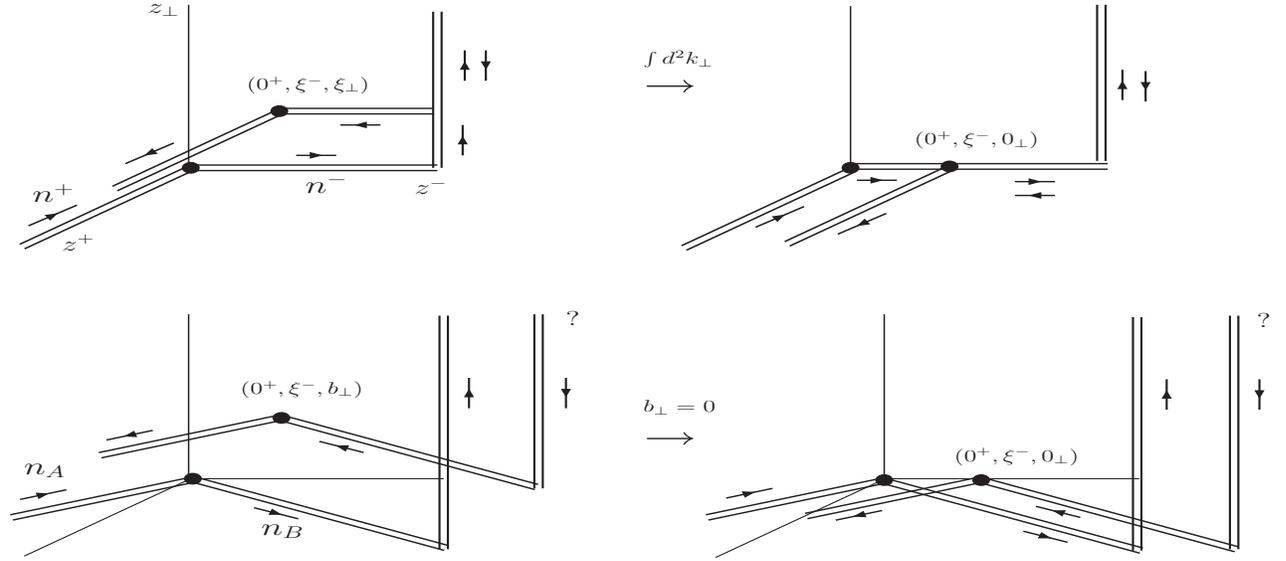}
\caption{\label{fig:geo2}Comparative layout of  Wilson lines in unsubtracted soft factors and visualization of the reduction to the collinear case. The upper panel shows the soft factor in momentum space, as proposed in Refs. \cite{CS_all}. The lower panel presents the tilted off-light-cone integration paths in impact parameter space, as well as the result of the reduction to the collinear $\bm b_\perp \to 0$ configuration.}
\end{figure}

\begin{figure}[p]
\includegraphics[width=0.45\textwidth,height=0.70\textheight,angle=90]{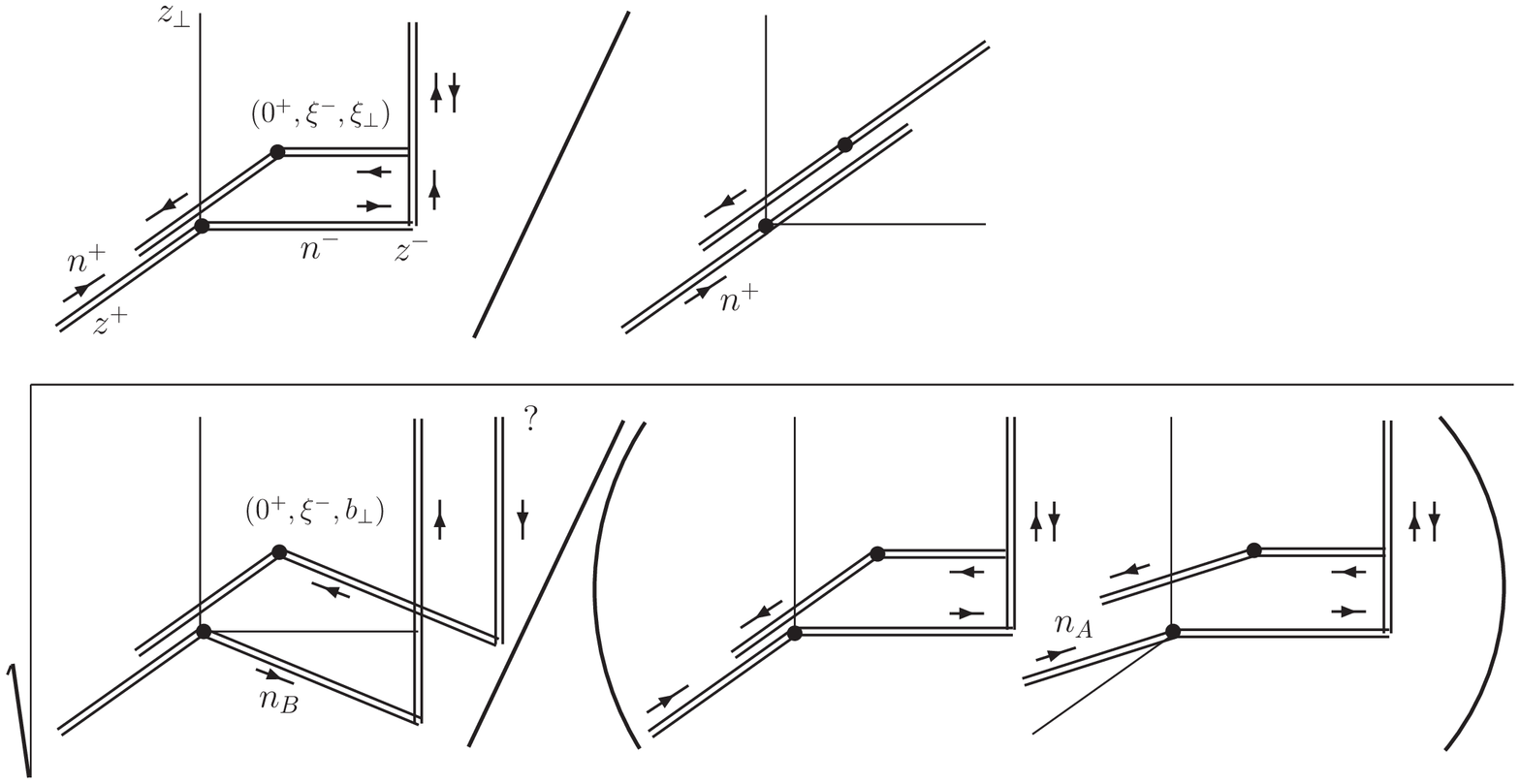}
\caption{\label{fig:geo3}Comparative layout of Wilson lines in subtracted soft factors. The upper panel corresponds to the soft factor of the TMD distribution function which enters the factorization with pure light-like Wilson lines. The lower panel {has the same setup, but with} the longitudinal Wilson lines shifted off the light-cone.}
\end{figure}

Let us discuss the emergent singularities arising in TMDs beyond tree-approximation. At one-loop level, the following three classes of divergences appear: $(i)$ standard ultraviolet poles, which are removable by a normal renormalization procedure; $(ii)$
pure rapidity divergences, which depend on an additional rapidity cutoff, but do not violate renormalizability of TMDs; they can be resummed by means of the Collins-Soper evolution equation; $(iii)$ very specific overlapping divergences: they contain the ultraviolet and rapidity poles simultaneously and thus break down the standard renormalizability of TMDs. This situation resembles the problems with renormalizability of the light-like Wilson loops discussed above.
However, the structure of Wilson lines is quite involved already in the tree-approximation.
The most straightforward definition of ``a quark in a quark'' TMD, which meets the requirement of the {parton number interpretation}, reads
\begin{eqnarray}
& & {\cal F}_{\rm unsub.} \left(x, {\bm k}_\perp \right)
=
  \frac{1}{2}
  \int \frac{d\xi^- d^2 {\xi}_\perp}{2\pi (2\pi)^2} \
  {\rm e}^{-ik \cdot \xi} \cdot \nonumber \\
& &   \times \left \langle
              p \ |\bar \psi_a (\xi^-,  \bm{\xi}_\perp)
              {\cal W}_{n}^\dagger(\xi^-,  \bm{\xi}_\perp;
   \infty^-,  \bm{\xi}_\perp) {\cal W}_{\bm l}^\dagger(\infty^-,  \bm{\xi}_\perp;
   \infty^-,  {\infty}_\perp) \cdot \right.  \nonumber \\
& &  \left.
 \times
   \gamma^+ {\cal W}_{\bm l}(\infty^-,  {\infty}_\perp;
   \infty^-, \bm{0}_\perp)_{\bm l}
   {\cal W}_{n}(\infty^-, \bm{0}_\perp; 0^-,\bm{0}_\perp)_{n}
   \psi_a (0^-,\bm{0}_\perp) | \ p \right \rangle \
\label{eq:general}
\end{eqnarray}
with ${\xi^+=0}$.
Here we define the semi-infinite Wilson lines evaluated along a four-vector $w$ as
$$
{\cal W}_w(\infty; \xi)
\equiv
  {\cal P} \exp \left[
                      - i g \int_0^\infty d\tau \ w_{\mu} \
                      A_{a}^{\mu}t^{a} (\xi + w \tau)
                \right] \ ,
$$
where the vector $w$ can be light-like $w_L = n^\pm\ , \ (n^\pm)^2 =0$, or transverse $w_T = {\bm l}$. Formally, the integration of (\ref{eq:general}) over $\bm k_\perp$ is expected to give the collinear (also called integrated) PDF
\begin{equation}
  \int\! d^2 \bm k_\perp \ {\cal F}_{\rm unsub.} (x, \bm k_\perp)
  =
 \frac{1}{2}
   \int \frac{d\xi^- }{2\pi } \
  {\rm e}^{-ik^{+}\xi^{-} } \
  \left\langle
              p\  |\bar \psi_a (\xi^-, \bm 0_\perp){\cal W}_n(\xi^-, 0^-)
              \gamma^+
   \psi_a (0^-,\bm 0_\perp) | \ p
   \right\rangle \ = f_a(x) \ .
   \label{eq:u_to_i}
\end{equation}
However, this is only justified in tree approximation.
It is worth noting that the normalization of the above TMD
\begin{equation}
{\cal F}_{\rm unsub.}^{(0)} (x, {\bm k}_\perp)
=
  \frac{1}{2}
  \int \frac{d\xi^- d^2
 \bm{\xi}_\perp}{2\pi (2\pi)^2}
  {\rm e}^{- i k^+ \xi^- + i  \bm{k}_\perp \cdot  \bm{\xi}_\perp}
 { \langle p \ | }\bar \psi (\xi^-,  \bm{\xi}_\perp)
   \gamma^+
   \psi (0^-, \bm 0_\perp) { | \ p \rangle } =
   \delta(1 - x ) \delta^{(2)} (\bm k_\perp) \
\label{eq:tree_tmd}
\end{equation}
can be most easily obtained by making use of the {canonical quantization procedure in the light-cone gauge}, where longitudinal Wilson lines become equal to unity and where equal-time commutation relations for creation and annihilation operators $\{a^\dag (k, \lambda), a(k, \lambda)\}$ immediately yield the parton number interpretation
\begin{equation}
 {\cal F}_{\rm unsub.}^{(0)} (x, {\bm k}_\perp)
 \sim
 \langle\  p \  | \ a^\dag(k^+, \bm k_\perp; \lambda) a(k^+, \bm k_\perp; \lambda) \ | \ p\ \rangle
 \ .
 \label{eq:parton_N}
\end{equation}
The usage of ``tilted'' gauge links in the operator definition of TMDs does not meet this requirement. We visualize the geometrical layout of various Wilson lines in the operator definition of TMDs in Fig. \ref{fig:geo1}, \ref{fig:geo2}, \ref{fig:geo3} and discuss relevant issues in their captions.


\begin{figure}[ht]
 $$\includegraphics[angle=90,scale=0.7]{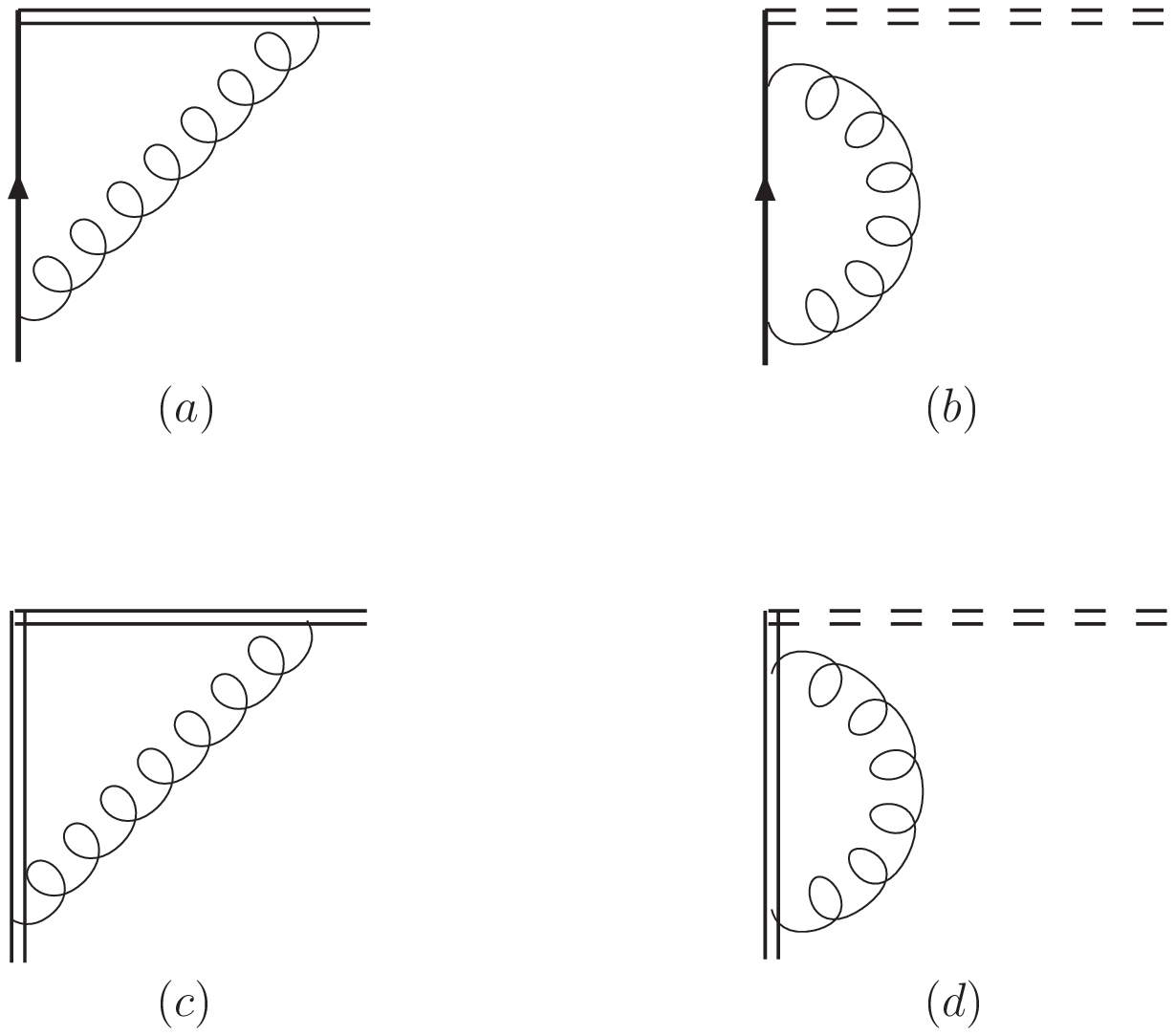}$$
   \vspace{0.0cm}
   \caption{\label{fig:1}The virtual one-loop Feynman graphs which produce extra singularities: $(a)$---vertex-type fermion-Wilson line interaction in covariant gauge; $(b)$---self-energy graph which yields the extra divergency in light-cone gauge; $(c,d)$ are the counter-parts of $(a,b)$ from the soft factor made of Wilson lines.}
\end{figure}

Beyond tree-approximation, the virtual diagrams producing terms with overlapping singularities are shown in Fig.\ \ref{fig:1}. The typical extra divergency stems from the one-loop vertex-type graph Fig. \ref{fig:1}(a) in covariant gauges or from the self-energy graph Fig. \ref{fig:1}(b) in the light-cone gauge (in the large-$N_c$ limit) and reads
\begin{equation}
{\rm TMD}_{\rm UV\otimes LC}
= - \frac{\alpha_s N_c}{2\pi}
 \Gamma (\epsilon) \left[ 4\pi \frac{\mu^2}{-p^2} \right]^{\epsilon}  \
 \delta (1-x) \delta^{(2)} (\bm{k}_\perp ) \ {\int_0^1\! dx \ \frac{x^{1-\epsilon}}{(1-x)^{1+\epsilon}} } \ .
\end{equation}
The standard ultraviolet pole in the Gamma-function $\Gamma (\epsilon)$ is accompanied by an additional singularity in the integral. The latter is due to the integration over infinite gluon rapidity and cannot be treated by dimensional regularization, calling for an extra (rapidity) cutoff. The reason for renormalizability violation in the leading order contribution to TMDs is that light-like Wilson lines (or the ``standard'' quark self-energy in light-cone gauge) produce more singular terms than usual Green functions do.

To solve the problems with extra singularities and renormalizability in TMDs, a variety of (possibly non-equivalent)  methods has been proposed. Working in the covariant Feynman gauge, Ji, Ma and Yuan proposed a scheme which utilizes tilted (off-light-cone) longitudinal Wilson likes directed along the vector $n_B^2 \neq 0$ \cite{JMY}. Transverse Wilson lines at the light-cone infinity cancel in covariant gauges, while the rapidity cutoff $\zeta = (2 p\cdot n_B)^2/|n_B^2|$ marks the deviation of longitudinal Wilson lines from a pure light-like direction. A subtracted soft factor then contains non-light-like Wilson lines as well. Obviously, such off-light-cone unsubtracted TMDs with the light-like vector $n^-$  replaced by the vector $n_{\rm B} = (-{\rm e}^{2y_B}, 1, \bm 0_\perp)$ do not obey the equation (\ref{eq:u_to_i}), not even {at tree level}. However, it is possible to formulate a ``secondary factorization'' method which allows one to express off-light-cone TMDs (in impact parameter space ${\cal F} (x, \bm b_\perp)$) as a convolution of integrated PDFs and perturbative coefficient functions in the perturbative region (that is, at small $\bm b_\perp$), see \cite{JMY}.

In publications \cite{CS_all} it was proposed to explore the renormalization-group properties of unsubtracted TMDs (\ref{eq:general}) and to make use of their anomalous dimension as a tool to discover the {minimal} layout of Wilson lines in the soft factor that provides a cancelation of overlapping dependent terms. It has been demonstrated (in the leading $O(\alpha_s)$-order) that the extra contribution to the anomalous dimension is exactly the cusp anomalous dimension \cite{KR87}, which is a crucial element of the investigation of non-renormalizible cusped light-like Wilson loops. Making use of specially chosen soft factors, one can get rid of the extra divergences in the operator definition of the TMDs, however paying a price in the form of significant complication of the structure of the Wilson lines in the above definition. In the present work we discuss another approach to the problems of light-cone cusped Wilson loops \cite{ChMVdV_2012}. To this end, it appears instructive to study those properties shared by such apparently different quantities as TMDs, light-like Wilson polygons, etc., which originate in their light-cone structure and arise in the form of the ``too singular'' non-renormalizable terms.

\section{Schwinger dynamical principle and area evolution for smooth Wilson loops}

We  made use of the observation that in the large-$N_c$ limit, in the transverse null-plane, for the light-like {planar} {dimensionally regularized (not renormalized)} Wilson rectangles, the area derivatives introduced in the previous sections can be reduced to the normal ones. The area variational equations in the coordinate representation describe the evolution of light-like Wilson polygons and represent, therefore, the ``equations of motion'' in loop space, valid for a specific class of its elements. As a result, the obtained differential equations give us a closed set of dynamical equations for the loop functionals, and can in principle be solved in several interesting cases.

Let us start with the {quantum dynamical principle} proposed by Schwinger \cite{Schwinger51}: the action operator $S$ defines  variations of arbitrary quantum states, so that
\begin{equation}
 {  \delta \langle \ \alpha' \ | \ \alpha''\  \rangle }
 =
 \frac{i}{\hbar} {  \langle \ \alpha' \ | \delta S | \ \alpha'' \ \rangle } \ .
 \label{eq:principle}
\end{equation}
The area variations (\ref{eq:area_derivative}) of field exponentials $\Phi(\Gamma)$  yield
\begin{equation}
 {  \frac{\delta}{\delta \sigma} \langle \ \alpha' \ | \Phi(\Gamma) | \ \alpha''\  \rangle }
 =
 \frac{i}{\hbar} {  \langle \ \alpha' \ | \frac{\delta \hat S}{\delta \sigma} \Phi(\Gamma) | \ \alpha'' \ \rangle } \ ,
 \label{eq:principle_mod}
\end{equation}
where $\hat S$ is yet to be defined.
The loop space consists of scalar objects with different topological and geometrical features, hence the equations of motion in this space must be the laws which state how those objects change their shape. It means that ``motion'' in loop space is equivalent to  variation of the shapes of integration contours in the Wilson loops \cite{MM_WL}. Therefore, we have to find the proper operator $\hat S$, which governs the shape variations of the light-like cusped loops (Wilson planar polygons).

Following the standard method, one makes use of Eq. (\ref{eq:principle}) in the form (\ref{eq:sch_dy_YM}) and gets the system of the MM Eqs. (\ref{eq:MM_general}). Alternatively, we will try to get rid of the operations which tacitly assume the property of smoothness of the Wilson loops of interest. Consider, e.g., a generic Wilson loop $W (\Gamma)$ without mentioning if it is smooth or not. The two leading terms of its perturbative series are given by
\begin{equation}
 {\cal W} (\Gamma) =
 {\cal W}^{(0)} + {\cal W}^{(1)} = 1 - \frac{g^2 C_F}{2} \ \oint_\Gamma \oint_\Gamma \ dz_\mu dz_\nu' \ D^{\mu\nu} (z - z') + O(g^4) \ , \nonumber
 \label{eq:W_generic_pert}
 \end{equation}
where $D^{\mu\nu}$ is the dimensionally regularized ($\omega = 4 - 2 \epsilon$) free gluon propagator
\begin{equation}
  D^{\mu\nu}
  =
  - g^{\mu\nu} \ \Delta (z - z') \ , \
  \Delta(z-z')
  =
  \frac{\Gamma(1-\epsilon)}{4\pi^2} \ \frac{(\pi \mu^2)^\epsilon}{[- (z-z')^2 + i0]^{1- \epsilon}} \ .
  \label{eq:gluon_prop_DR}
\end{equation}
For convenience's sake, we work in the Feynman covariant gauge and separate out the scalar part of the propagator $\Delta (z)$. The issues related to  gauge- and regularization independence of the calculations will be considered elsewhere. Then, if the l.h.s. of the Eq. (\ref{eq:principle_mod}) acts on the Wilson exponential (\ref{eq:W_generic_pert}), one gets
\begin{equation}
 \frac{\delta {\cal W} (\Gamma)}{\delta \sigma_{\mu\nu}}  =
 \frac{g^2 C_F}{2} \ \frac{\delta }{\delta \sigma_{\mu\nu}}  \oint_\Gamma \oint_\Gamma \ dz_\lambda dz^{'\lambda} \ \Delta (z - z') + O(g^4) \ .
 \label{eq:W_area_diff_2}
 \end{equation}
Using the Stokes theorem (let us assume for a moment that we are allowed to do so), we obtain
\begin{equation}
 \oint_\Gamma dz_\lambda \ {\cal O}^\lambda
 =
 \frac{1}{2} \int_\Sigma \ d\sigma_{\lambda \rho} (\partial^\lambda {\cal O}^\rho - \partial^\rho {\cal O}^\lambda) \ , \
{\cal O}^\lambda
 =
 \oint_\Gamma dz_\lambda' \ \Delta (z) \ ,
\label{eq:W_diff_Stokes}
\end{equation}
where $\Gamma$ is considered as the boundary of the surface $\Sigma$.
We get then the leading perturbative term of the MM equation (\ref{eq:MM_general}):
\begin{equation}
 \partial_\mu \frac{\delta {\cal W}(\Gamma_\bigcirc)  }{\delta \sigma_{\mu\nu} (x)}
 =
 \frac{g^2N_c}{2} \oint_{\Gamma_\bigcirc} \ dy_\nu \ \delta^{(\omega)} (x -y) + O(g^4)\ .
 \label{eq:MM_LO_smooth}
\end{equation}
We must treat this result with due caution: in the course of the derivation, we assumed that the Stokes theorem is valid for all Wilson loops which we consider. However, the last statement is not true in general, that is why we mark the ``good'' contours with a special index $\Gamma_\bigcirc$. It is interesting that in $2D$ QCD, the area differentiation turns into the ordinary derivative, by virtue that the gluon propagator (\ref{eq:gluon_prop_DR}) for $\omega=2$ gets logarithmic in $z$:
\begin{equation}
  {\cal W}(\Gamma_\bigcirc)^{\rm 2D}
  =
  \exp\[ - \frac{g^2 N_c}{2} \Sigma \] \ , \ \Sigma = \ {\rm area \ inside\ } \Gamma_\bigcirc \ ,
  \label{eq:2D_area}
\end{equation}
so that
$
  2 \ln W'_\Sigma =
  - {g^2 N_c}.
$
Evaluating, in the same way, the NLO terms, one obtains the full MM Eq. (\ref{eq:MM_general}). However, we interrupt here and return a bit backward, since we are mostly interested in those Wilson functionals that do not satisfy (or, at least, do not satisfy straightforwardly) the conditions of the applicability of the Stokes theorem. We will, therefore, continue with the study of the shape variations of Wilson loops without relying upon the Stokes theorem, but keeping in mind an explicit form of the free gluon propagator (which possesses a specific light-cone/rapidity divergence), see Eq. (\ref{eq:gluon_prop_DR}).


\section{Singularities of Wilson rectangles}

We are now in a position to extend the Schwinger approach to a more complicated case and to try to derive the corresponding area evolution equations. The calculation of cusped light-cone Wilson loops beyond tree approximation in different  gauges and the justification of gauge independence calls for a careful treatment of a variety of divergences already in leading order. Special attention must by paid to the separation of the rapidity divergences and the standard ultraviolet ones \cite{WL_RG, WL_LC_rect, WL_LC_rapidity}.
In the 't Hooft (large-$N_c$) limit one obtains \cite{WL_LC_rect}
\begin{eqnarray}
& & W(\Gamma_\Box)  = 1 - \frac{1}{\epsilon^2}\ \frac{\alpha_s N_c}{2\pi} \ \(\[{-2 N^+ N^-\mu^2 + i0}\]^\epsilon + \[{2 N^+ N^-\mu^2 + i0}\]^\epsilon  \) \\ & &  + \frac{\alpha_s N_c}{2\pi} \( \frac{1}{2} \ln^2 \frac{N^+N^-+ i0}{-N^+N^-+i0} + {\rm finite\ terms}  \) + O(\alpha_s^2) \ , \nonumber
\label{eq:WL_LC_1loop}
\end{eqnarray}
with the Mandelstam variables in the momentum space, ${ s = (p_1 + p_2)^2}$ and ${t = (p_2 + p_3)^2}$, map onto the {area} variables in the coordinate space, so that ${ s/2} = {- t/2} \to N^+ N^-$.
We will show separately that the result (\ref{eq:WL_LC_1loop}) is not only gauge invariant, but is independent of any regularization of light-cone and ultraviolet divergences and of the way they are separated. The latter issue is of considerable importance to the study of the operator structure of transverse-momentum dependent parton densities, the jet and soft functions in the soft-collinear effective theory, the infrared properties of the high-energy scattering amplitudes, etc. (see, e.g., \cite{CS_all, SCET_TMD} and references therein).

\begin{figure}[ht]
 $$\includegraphics[angle=90,width=0.7\textwidth]{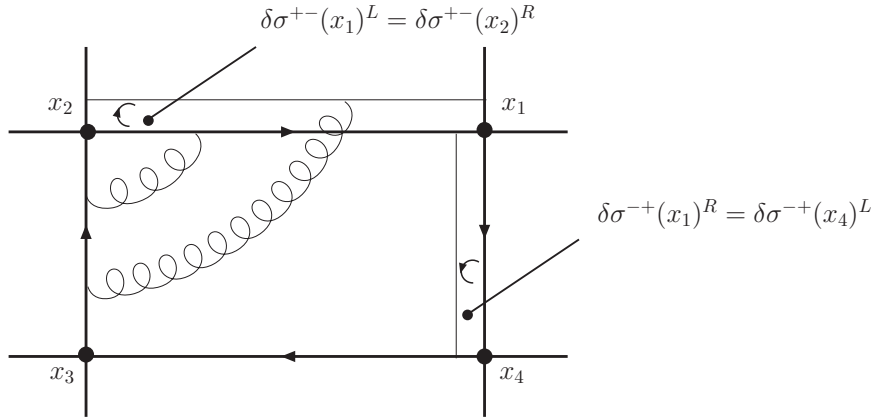}$$
   \caption{Infinitesimal area transformations for a light-cone rectangle on the null-plane: we consider only those area variations that conserve the angles between the sides. These variations are defined at the corners $x_i$.}
   \label{fig:2}
\end{figure}

The transverse null-plane is defined by the condition $\vecc z_\perp = 0$; therefore, the shape variations (which give rise to the infinitesimal area changes) are defined as
\begin{equation}
 { \delta \sigma^{+-} }
=
{ N^+ \delta N^- }  \ ,  \
{ \delta \sigma^{-+} }
=
- { N^- \delta N^+ }  \ .
\label{eq:delta_area}
\end{equation}
We assume that these operations do make sense only at the corner points ${x_i}$, and introduce the ``left'' and ``right'' variations, as shown in Fig. \ref{fig:2}.

Rectangular planar Wilson loop ${ W(\Gamma_{\Box})}$ is known to lack multiplicative renormalizability \cite{WL_LC_rect}.  In order to decrease the power of singularity that violates the renormalizability, one can follow the scheme proposed in \cite{CAD_universal}. Having in mind Eq. (\ref{eq:delta_area}), we define the area logarithmic derivative
\begin{equation}
  \frac{\delta}{\delta \ln \sigma}
  \equiv
  \sigma_{+-} \frac{\delta}{\delta \sigma_{+-}}
  +
  \sigma_{-+} \frac{\delta}{\delta \sigma_{-+}} \
  \label{eq:area_log}
\end{equation}
and act with this operator on the r.h.s. of the Eq. (\ref{eq:WL_LC_1loop}):
\begin{equation}
\frac{\delta}{\delta \ln \sigma} \ \ln W(\Gamma_\Box)
= - { \frac{\alpha_s N_c}{2\pi} } \ \frac{1}{\epsilon} \ \( \[{-2N^+N^-\mu^2  + i0}\]^\epsilon + \[{2N^+N^-\mu^2 + i0}\]^\epsilon \) \ .
\label{eq:log_der}
\end{equation}
Then, the result is finite (after additional logarithmic differentiation in the ultraviolet scale $\mu$) and is given by the {cusp anomalous dimension}
\begin{equation}
 \mu \frac{d}{d\mu} \frac{\delta \ \ln W(\Gamma_\Box)}{\delta \ln \sigma}
 =
 - 4 \ { \Gamma_{\rm cusp} } \ , \ {\Gamma_{\rm cusp} = \frac{\alpha_s N_c}{2 \pi} } + O(\alpha_s^2) \ .
 \label{eq:full_der}
\end{equation}
Therefore, we have obtained the result (\ref{eq:full_der}) given that the infinitesimal shape variations are defined as in (\ref{eq:area_derivative}). Eq.(\ref{eq:full_der}) describes then the dynamics of the cusped planar light-like Wilson loops \cite{ChMVdV_2012}. We established, therefore, the connection between the {geometry} of the loop space (in terms of the area/shape differentials) and the {dynamical properties} of the fundamental degrees of freedom---the gauge- and regularization-invariant planar {light-like Wilson loops}.

\section{Combined evolution from the Schwinger principle}

The very possibility to obtain a finite result by means of Eqs. (\ref{eq:log_der}, \ref{eq:full_der}) is a direct consequence of the geometrical properties of loop space, when considering the non-renormalizable cusped light-like Wilson loops.
To show this explicitly, we restrict ourselves to shape variations  (\ref{eq:delta_area}), and apply the area derivative to the planar Wilson rectangle
\begin{eqnarray}
 & & \frac{\delta W(\Gamma_\Box)  }{\delta \sigma_{\mu\nu}} = \frac{g^2C_F}{2} \frac{\Gamma(1-\epsilon) (\pi \mu^2)^\epsilon}{4\pi^2} \
 \frac{\delta }{\delta \sigma_{\mu\nu}} \sum_{i,j} (v_j^\lambda v_j^\lambda) \cdot \nonumber  \\
 & & \times \int_0^1\int_0^1\! \frac{d\tau d\tau'}{[- (x_i - x_j - \tau_i v_i + \tau_j v_j)^2 + i0]^{1 -\epsilon}} \ ,
 \label{eq:WL_rect_area}
\end{eqnarray}
where the sides are parameterized as $z_i^\mu = x_i^\mu - v_i^\mu \tau$ with the vectors $v_i$ having the dimension $[{\rm mass}^{-1}]$ \cite{WL_LC_rect}.
A peculiar feature of the planar light-like contours is that the area dependence separates out from the integrals and can be calculated explicitly (taking into account that $2 (v_1v_2) = 2 N^+N^-$, see Eq. (\ref{eq:delta_area}))
\begin{equation}
W^{(1)}(\Gamma_\Box)
=
- \frac{\alpha_s N_c}{2\pi} {\Gamma(1-\epsilon)} (\pi \mu^2)^\epsilon
\ (- 2 N^+N^-)^\epsilon\ \frac{1}{2} \int_0^1\int_0^1\! \frac{d\tau d\tau'}{[(1-\tau)\tau']^{1 -\epsilon}}   \ .
 \label{eq:WL_rect_area_NN}
\end{equation}
Moreover, light-like Wilson lines with $v_i^2 = 0$ develop an extra singularity, which shows up in the form of a second-order pole  $\sim \epsilon^{-2}$, while the cusps violate conformal invariance of the Wilson loop because the ``skewed'' scalar products $(v_i v_j) \neq 0$ replace the conformal ones $v_i^2$. Hence, performing the area
$\delta / \delta \ln \sigma = \delta / \delta \ln (2 N^+N^-) $
and the UV-scale logarithmic differentiation of Eq. (\ref{eq:WL_rect_area_NN}) and summing up all relevant terms, we obtain
\begin{equation}
 \mu \frac{d }{d \mu } \ {\[  \frac{\delta}{\delta \ln \sigma} \ \ln \ W (\Gamma) \] }
 =
 - \sum { \Gamma_{\rm cusp} } \ ,
 \label{eq:mod_schwinger}
\end{equation}
which has been foreseen in Eq. (\ref{eq:full_der}) and which is derived now directly from the Schwinger quantum dynamical principle. It is natural that this result is akin, in some sense, to the situation in $2D$ QCD mentioned above. The area derivative becomes the ordinary derivative for the same reason: one has effectively planar Wilson loops, thus the MM Eqs. becomes a closed system and, in principle, solvable \cite{MM_WL, WL_Renorm}.

Let us emphasize that the r.h.s. of Eq. (\ref{eq:mod_schwinger}) is regulated by the cusp anomalous dimension, which is a universal quantity that is independent of the shape of particular contour and which is known perturbatively up to the $O(\alpha_s^3)$ order.  It is therefore worth analyzing if the above result is only a leading order approximation, or if it also will be valid for higher order approximations. Let us take into consideration the property of linearity of the (angular-dependent) cusp anomalous dimension in the infinitely large angle asymptotics with respect to the logarithm of the cusp angle $\chi \to  \frac{1}{2}\ln \frac{(2 v_i v_j)^2}{v_i^2 v_j^2}$ \cite{KR87}:
\begin{equation}
 \lim_{\chi \to \infty} \Gamma_{\rm cusp} (\chi, \alpha_s)
 =
 \sum \alpha_s^n C_n (W) a_n (W) \ \ln \frac{(2 v_i v_j)}{|v_i| |v_j^2|} \ ,
 \label{eq:cuspCD_LC}
\end{equation}
with the ``maximally non-Abelian'' coefficients given by
\begin{equation}
 {C_k} \sim C_F \ N_c^{k-1} \to {\frac{N_c^k}{2} } \ ,
\end{equation}
and $a_n$ are cusp-independent factors.
This asymptotic regime coincides with the light-cone case with the angular-dependent logarithms being transformed into extra pole terms in $\epsilon$: $\chi \to \frac{(v_iv_j)^\epsilon}{\epsilon}$, see \cite{WL_LC_rect, KR87}.
More specifically, the area variable $ \ \sim (v_iv_j)$ enters the regularized cusp anomalous dimension in the light-like limit as
\begin{equation}
 \Gamma_{\rm cusp} ({\rm area}, \epsilon, \alpha_s)
 =
 \sum \alpha_s^n C_n (W) a_n (W) \ \frac{{\rm area}^\epsilon}{\epsilon} \ ,
 \label{eq:cuspCD_LC_1}
\end{equation}
and, after logarithmic differentiation, one obtains a perturbative expansion of the cusp anomalous dimension
\begin{equation}
 \lim_{\epsilon \to 0} \frac{d \ \Gamma_{\rm cusp} ({\rm area}, \epsilon, \alpha_s)}{d\ \ln {\rm area}}
 =
 \sum \alpha_s^n C_n (W) a_n (W) \ ,
 \label{eq:cuspCD_LC_2}
\end{equation}
which justifies the previous result (\ref{eq:mod_schwinger}) in the NLO by virtue that
$$
\Gamma_{\rm cusp} = - d \ln W / d \ln \mu \ .
$$
This implies that the result (\ref{eq:mod_schwinger}) is, in fact, an all-order one, akin to the MM Eq. (\ref{eq:MM_general}): they both are exact and non-perturbative, while the r.h.s's of each one can be calculated perturbatively.
It is worth noting that Eq. (\ref{eq:full_der}) agrees with the non-Abelian exponentiation theorem for the dimensionally regularized Wilson loops
\begin{equation}
 W (\Gamma_\Box; \epsilon)
 =
 \exp \[ \sum_{k=1} \alpha_s^k \ C_k (W) F_k (W) \ \] \ ,
\end{equation}
with the summation going over all two-particle irreducible diagrams, whose contribution is given by the so-called ``web'' functions $F_k$ \cite{WL_expo, KR87}.
Thus, Eq. (\ref{eq:full_der}) can be used in calculation of the higher-order terms in the cusp anomalous dimension, by virtue that one has a closed recursive system of the perturbative equations.

\begin{figure}[ht]
 $$\includegraphics[angle=90,scale=0.7]{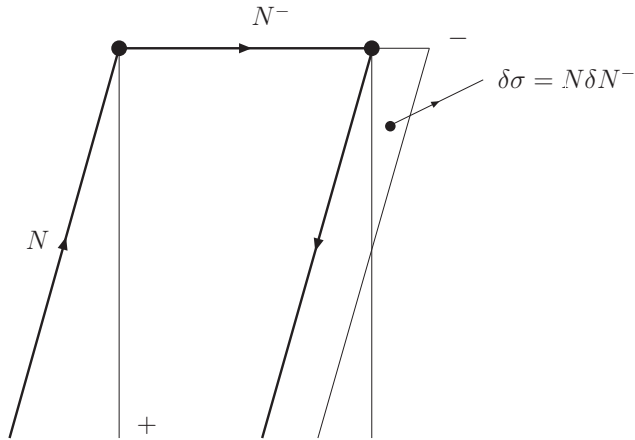}$$
   \vspace{0.2cm}
   \caption{\label{fig:5}$\Pi$-shape integration contour and the infinitesimal area variations.}
\end{figure}

\begin{figure}[ht]
 $$\includegraphics[angle=90,width=0.7\textwidth]{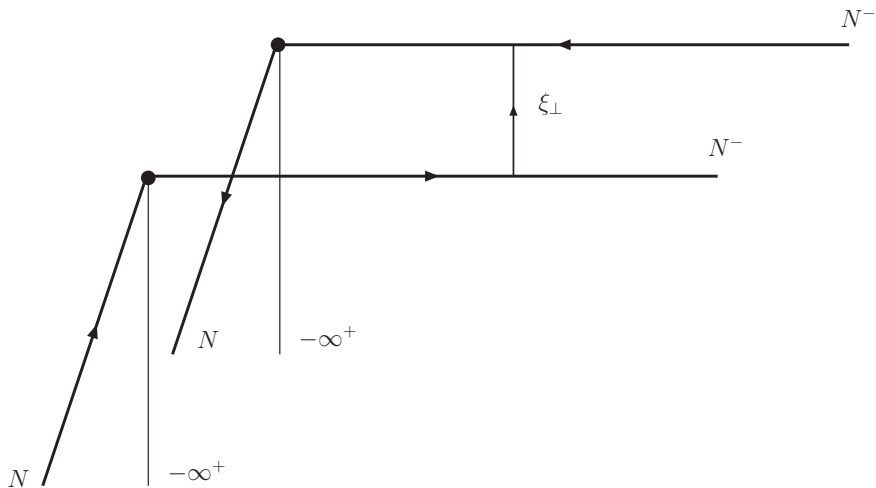}$$
   \caption{Conjectured ``dual'' Wilson loop having the combined evolution similar to the one of a TMD. Transverse Wilson lines are not shown for simplicity.}
   \label{fig:7}
\end{figure}

Now we can apply the methodology described above to the TMDs with the light-like longitudinal Wilson lines ${\cal F} (x, \vecc k_\perp)$, Eq. (\ref{eq:general}), what yields
\begin{equation}
  \mu \frac{d }{d \mu } \ \[ \frac{d}{d \ln \theta} \ \ln \ { {\cal F} (x, \vecc k_\perp) } \]
  =
  2 { \Gamma_{\rm cusp} } \ ,
  \label{eq:tmd_combined}
\end{equation}
with the ``area'' being encoded in the rapidity cutoff parameter $\theta \sim (p N^-)^{-1}$ \cite{CS_all}. Another interesting example is provided by the $\Pi$-shape semi-loop with one finite light-like segment \cite{WL_Pi}. In the one-loop order, we have in the large-$N_c$ limit
\begin{eqnarray}
  & & W(\Gamma_\Pi)
  =
  1 + \frac{\alpha_s N_c}{2\pi} \  +
  \[ - L^2 (NN^-) + L (NN^-)  - \frac{5 \pi^2}{24}  \] \ , \\
  & & L(NN^-)
  = \frac{1}{2}\(\ln (\mu N N^- + i0) + \ln (\mu NN^- + i0) \)^2 \ , \nonumber
  \label{eq:pi_1loop}
  \end{eqnarray}
where the area is given by the product of the light-like $N^-$ and non-light-like $N$ vectors, see Fig. \ref{fig:5}.
The $\Pi$-shaped Wilson loop (\ref{eq:pi_1loop}) also satisfies Eq. (\ref{eq:mod_schwinger}):
\begin{equation}
  \mu \frac{d }{d \mu } \ \[ \frac{d}{d \ln \sigma} \ \ln \ { W(\Gamma_\Pi) } \]
  =
  - 2 { \Gamma_{\rm cusp} } \ ,
\end{equation}
the latter controls for the renormalization-group evolution of the integrated PDFs in the large-$x$ limit, as well as the anomalous dimensions of conformal operators with large Lorentz spin \cite{WL_Pi}. The $\Pi$-shape contour can be split and moved apart to separate two planes by the transverse distance $\bm \xi_\perp$. The Wilson loop obtained in such a way is expected to be ``dual'' to the TMD. This duality implies that both the double-planar $\Pi$-shaped Wilson loop, Fig. \ref{fig:7}, and the TMD (\ref{eq:general}) obey the same combined evolution equation (\ref{eq:tmd_combined}). The detailed analysis of this configuration will be presented elsewhere.

\section{Conclusions and outlook}
The universal quantum dynamical principle by Schwinger  provides a proper approach to the description of the dynamics of the loop space. The gauge invariant Wilson loops are considered then as the only degrees of freedom, and the Makeenko-Migdal equations (\ref{eq:MM_general}) stem from the Schwinger-Dyson equations applied to the renormalizable loops. In general case, the system of the MM Eqs. is not closed and cannot be immediately used in practical calculations in QCD.

\begin{figure}[ht]
 $$\includegraphics[angle=90,scale=0.7]{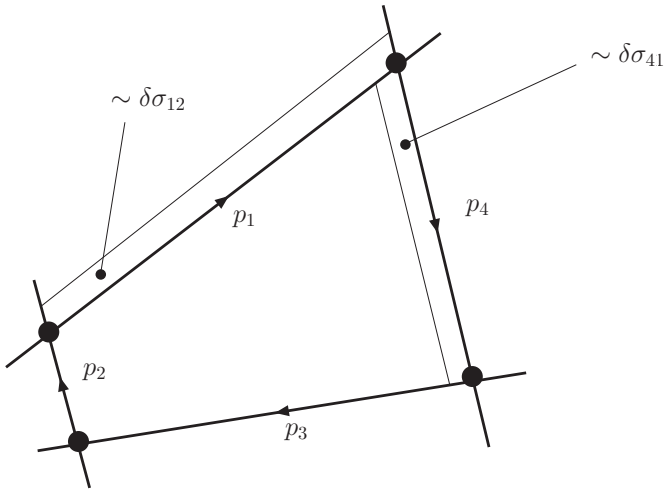}$$
   \caption{\label{fig:3}Generic infinitesimal area variations responsible for the conjectured quantum-dynamical loop equations for light-like Wilson $n-$polygons. Evaluation of minimal surface differentials for more complicated cusped Wilson loops is required  to derive corresponding area evolution equations based on the quantum dynamical principle \cite{WL_min_surf}.}
\end{figure}

In the present work we showed that it is possible to design a relevant system of equations of motion valid for the cusped planar light-like Wilson loops, taking into account that the latter possess a very specific singularity structure compared to their off-light-cone analogues. General solution of this problem has not been found yet, but we have managed to demonstrate that some simplifications make it possible to propose a new potentially fruitful method to deal with such Wilson exponentials.
Namely, in the large-$N_c$ limit, the {planar} rectangular light-like contours at $\vecc z_\perp =0$ enable us to reduce the area functional derivative to the normal derivative for dimensionally regularized Wilson loops. As the result, the equations which describe the infinitesimal shape variations in coordinate space appear to be dual to the energy evolution equations for cusped Wilson loops in the space of the Mandelstam momentum variables.
Within the framework we proposed, the dynamics of elements of loop space is introduced by means of obstructions of the initially smooth Wilson loops, which play, therefore, the role of {\it sources} in Schwinger's ``fields and sources'' picture. We have argued, therefore, that the Schwinger quantum dynamical principle can be used as an effective tool to study (at least) one important class of the elements of loop space---the cusped planar Wilson polygons on the light-cone. We implemented the program only in one of the simplest situations, the planar rectangle. In Fig. \ref{fig:3}, a more involving configuration is visualized, the arbitrary quadrilateral path, the area evolution of which is far from being trivial and deserves a separate study.


\section{Acknowledgements}
We appreciate stimulating discussions and useful critical remarks by I.V. Anikin, Y.M. Makeenko and N.G. Stefanis. We thank the participants of the JLab QCD Evolution workshop, the summer meetings in ECT*, Trento, and the joint seminar at the University of Antwerp for useful discussions.


\end{document}